# Electrical spin injection and detection in lateral all-semiconductor devices.

M. Ciorga, A. Einwanger, U. Wurstbauer, D. Schuh, W. Wegscheider, and D. Weiss

University of Regensburg, Institute of Experimental and Applied Physics, Universitätsstrasse 31, D-93053 Regensburg, Germany

Abstract

Electrical injection and detection of spin-polarized electrons is demonstrated for the first time in a single wafer, all-semiconductor, GaAs-based lateral spintronic device, employing $p^+$-(Ga,Mn)As/$n^+$-GaAs ferromagnetic Esaki diodes as spin aligning contacts. The conversion of spin-polarized holes into spin-polarized electrons via Esaki tunnelling leaves its mark in a bias dependence of the spin-injection efficiency, which at maximum reaches the relatively high value of 50%.

The ability to inject, manipulate and detect spin-polarized carriers by purely electrical means is at the very heart of semiconductor spintronics [1,2]. Despite big progress on spin injection [2–11] into semiconductors, an all-semiconductor all-electrical injection and detection scheme remained so far elusive. The most successful concepts for lateral spin transport devices so far employed hybrid metal/semiconductor contacts to inject and detect spin-polarized carriers [10–11].

A discovery of ferromagnetism in semiconducting (Ga,Mn)As [12]) has provided the possibility to employ this material as an injector of spin-polarized carriers into a non-magnetic semiconductor in a single all-semiconductor device. The main advantage of this approach would be the compatibility of possible future spintronic devices with the existing technology of III-V semiconductors. The biggest disadvantage of (Ga,Mn)As as a spin injector is its *p*-type character resulting in short spin relaxation times because of the pronounced spin-orbit coupling in the valence band [13]. This obstacle has been recently overcome and the injection of spin-polarized electrons from (Ga,Mn)As into GaAs was detected optically using a $p^+$-(Ga,Mn)As/ $n^+$-GaAs Esaki diode structure as the injector contact [9,14,15]. Due to the high doping on both sides of such a structure, the top of the (Ga,Mn)As valence band overlaps energetically with the bottom of the GaAs conduction band (see Fig. 1b) and under a small reverse (forward) bias electrons from GaMnAs (GaAs) tunnel to GaAs (GaMnAs) [16] leading to spin injection (extraction).

We employ such diodes as both injecting and detecting contacts in our lateral device. The device operates in a non-local configuration [17] i.e. with a detector placed outside the current path, to minimize various spurious effects that could influence the measured signal. We verify the existence of a spin transport in GaAs channel by measuring the spin-valve effect [10,11,18,19] and Hanle effect [10,11,19,20] i.e., respectively, the switching in a non-local signal induced by the in-plane magnetic field and the oscillation and suppression of the non-local signal induced by a transverse magnetic field. We obtain relatively high spin polarization value of 50%, which is strongly effected by an applied bias.

A scanning electron micrograph (SEM) of the sample identical to the one measured is shown in Fig. 1(a). The device was fabricated from a single molecular-beam-epitaxy (MBE)-grown wafer using standard photolithography and etching techniques. The wafer, grown on a semi-insulating (001) GaAs substrate, consists of the following layers (in the order of growth): 500 nm of GaAs/AlGaAs superlattice, 250 nm of lightly Si-doped $n$-GaAs epilayer ($n=6\times10^{16}$ cm$^{-3}$), 15 nm of $n\rightarrow n^+$ Si-doped GaAs transition layer ($n^+=6\times10^{18}$ cm$^{-3}$), 8 nm of $n^+$-GaAs, 2.2 nm of low-temperature (LT)-grown Al$_{0.36}$Ga$_{0.64}$As and 20 nm of LT-grown Ga$_{0.95}$Mn$_{0.05}$As. Curie temperature of as-grown (Ga,Mn)As layer is ~65K, as established by magnetic measurements. LT-grown (Al,Ga)As layer is used to prevent a diffusion of Mn into GaAs. The tunnelling Esaki diode structure is formed by the $p$-type (Ga,Mn)As layer and $n^+$-GaAs layer. Current voltage (I-V) characteristics of the ferromagnetic contacts (2–5) confirm the presence of Esaki tunnelling (Fig.1(c)). For all measurements discussed here a dc current $I_{21}$ flows between contacts 2 and 1 and non-local voltages $V_{36}$ $V_{46}$, and $V_{56}$ are measured between respective contacts. According to the spin-injection theory [3,17,21], the non-local voltage is a measure of a non-equilibrium spin accumulation, induced in $n$-GaAs underneath the injector and diffusing in either direction of this contact. At a distance $L$ from the injector it holds

$$V^{nl} = \pm(P_{inj}P_{det}I\lambda_{sf}\rho_N/2S)\exp(-L/\lambda_{sf}),  \quad (1)$$

where $I$ is a bias current, $\rho_N$, $\lambda_{sf}$, $S$ are, respectively, resistivity, spin diffusion length and the cross-section area of the non-magnetic channel. P$_{inj(det)}$ is the spin injection efficiency of the injector (detector) contact and expresses the polarization of the current injected at the respective contact. $+(-)$ sign corresponds to a parallel (anti-parallel) configuration of magnetizations in ferromagnetic electrodes that can be switched by an in-plane magnetic field, as is done in a spin-valve (SV) experiment.

Typical results of SV experiments on our samples are presented in Fig. 2(a). The raw data is shown, which is a sum of a spin-related signal expressed by Eq. 1 and some background signal, observed in most non-local SV experiments [10,11,18,19], origin of which is still not well understood. The background signal shows some slight magnetic field dependence, which however can be neglected in the shown range of the magnetic field. A clear SV-like feature is observed for all three detectors. The amplitude ΔV of this feature decays exponentially with the injector-detector separation in a full agreement with Eq. (1), as is shown in the Fig. 2(b). All curves show also a sharp feature at 0T. Its dependence on the **B**-field sweeping rate suggests that it could be related to dynamic nuclear polarization (DNP) of GaAs nuclei [22] due to hyperfine interaction with the spins of injected electrons [23] (similar feature was also observed by Lou *et al.* [10]).

A further rigorous test of the system for spin-transport is a demonstration of the Hanle effect, i.e. the precession and dephasing of the injected spins during transport between injector and detector in magnetic field **B** perpendicular to their initial orientation [17,20,21]. The non-local voltage measured for parallel configuration of injector and detector can be expressed by [20,21]

$$V_{\parallel}(x_1, x_2, B) = V_0 \int_0^{\infty} \frac{1}{\sqrt{4\pi D t}} e^{-(x_2 - x_1 - v_d t)^2 / 4Dt} \cos(\Omega t) e^{-t/\tau_s} dt \quad (2)$$

where $V_0 = \pm P_{inj} P_{det} I \lambda_{sf} \rho_N / 2S$, $x_1$ and $x_2$ are respectively the point of injection and detection, $D$ is the spin diffusion constant, $\tau_s$ is the spin relaxation time, and $\Omega = g\mu_B B/\hbar$ is the precession frequency.

The typical results of Hanle experiments on our samples are shown in Fig. 2(c,d). The data was combined from two separate sweeps from $B_z=0$ in either direction of $B_z$, as sweeping through zero resulted in very asymmetric data, showing also hysteretic behaviour, that could be related to DNP effects [22]. Fig. 2(c) shows the raw data obtained at the detector 4 ($L$=10μm), for $I_{21}$=−50μA. As for the in-plane case this data is a superposition of the spin signal, described by Eq. 2, and the background. At low fields strong oscillations of the spin-related signal due to Hanle effect are clearly observed. The signal decreases from its maximum value (point I), changes sign when the average spin obtains a component anti-parallel to the magnetization direction of the detector (point II) and finally gets fully suppressed when all spins are dephased by $B_z$ (region III–IV). At B ~ 0.07T magnetization of (Ga,Mn)As contacts is rotated out of the sample plane and aligned along $B_z$ and the step-like feature (region IV–V) is observed in the data. At exactly this range of $B_z$ a step in the resistance of the injector circuit is also observed (not shown), which can be attributed to perpendicular-to-plane tunnelling anisotropic magnetoresistance (TAMR) [24−26]. The difference in signal between position V (out-of-plane magnetization) and I is then a measure of tunnelling anisotropic spin polarization (TASP). At higher fields (range V – VI) the spin signal is saturated and measured voltage follows the background, which, similarly as in work by Lou *et al* [10], has a parabolic dependence on $B_z$.

In Fig. 2(d) we show pure Hanle-related signal at all three detectors with the background removed. The signal is clearly symmetric in $B_z$ suggesting that the magnetizations of injector and detector are parallel $(\uparrow\uparrow)$ [20,21]. In agreement with theory more oscillations are visible for increased injector – detector distance. Model curves, based on Eq. 2, are also plotted in the figure as solid red lines. Using Eq. 2 we were able to estimate the spin polarization as $P$~0.4 and $\tau_s$, $\lambda_{sf}$ as respectively ~4 ns and ~3 μm, with the latter value being consistent with SV measurements. In agreement with the model the width of the Hanle curves increases with temperature, primarily due to the decrease in $\tau_s$ to 1.69 ns at 30K (not shown). There is a small discrepancy between the model and the measured data near 0T, where the linewidth of the measured signal is smaller than expected from the model. This also could be caused by hyperfine effects that can narrow the Hanle signal around zero magnetic field [22]. This discrepancy gets smaller when we move with the detector away from the injector, lower the bias current or increase the temperature.

Let us now shortly discuss the origin of the SV signal in our devices. As shape anisotropy of (Ga,Mn)As is considered to be very weak [27], it cannot be employed to switch between parallel and anti-parallel configuration of magnetizations. Magnetic equivalence of [010] and [100] directions, i.e. short and long axis of our ferromagnetic contacts is clearly seen in our data, as in-plane sweeps along both those directions give the same results (not shown). Some conclusions regarding the origin of the SV signal can be drawn from comparison SV data with the results of Hanle measurements. From the Hanle curves we know that in sufficiently high $B_z$ (region III–IV at Fig. 2(c)) all spins are fully dephased, the related spin signal is equal to zero and only background-related voltage is measured. For all our measurements its value is very close to the one measured at the top of the SV feature, i.e. at $B_x \approx 0.02T$ for traces at Fig. 2(a) (see also Fig. 3(a)). Therefore, we deduce that the spin signal at this value of $B_x$ is equal to zero (the $B$-dependence of the background is negligible in the relevant range). This can be confirmed by comparing both Hanle and SV curves taken at opposite bias $I$. According to Eq. 1 reversing $I$ also reverses the sign of the measured spin signal. One expects then that the signals measured for opposite $I$ will be mirror images of each other in respect to zero spin signal. As we can see in Fig. 3(a) curves taken for negative bias are almost mirror images of those for positive bias in respect to $V \approx 3.5 \mu V$ — the value measured at the top (bottom for negative bias) of the SV feature. As the detector reads the projection of the spin component on its magnetization axis this suggests that in this range of $B_x$ the spin-polarization is perpendicular to the magnetization of the detector. The SV feature observed during the sweep of $B_x$ could therefore be a result of a switching between parallel $(\uparrow\uparrow)$ and perpendicular $(\uparrow\leftarrow)$ configuration of the magnetizations of injector and detector. Although it would be very interesting to know the exact mechanism of this switching, it is however beyond the scope of our paper. We would like to mention only that the field position of the spin-valve feature coincides with the spin-valve-like feature observed in the resistance measured at the injector circuit (not shown). As the latter is the result of TAMR effect at the Esaki diode interface [26], this suggests that the observed perpendicular configuration is driven by magnetic reversal in the injector contact, influenced by the bias voltage.

Let us now discuss the actual value of the spin injection efficiency $P_{inj}$ in our devices. From model Hanle curves we extracted the value of $P$, equal to $P_{inj}$ when $P_{inj} = P_{det}$. The last equation, however, is generally not true as $P_{inj}$ is strongly bias-dependent. For all used bias values $\Delta V = V_{\uparrow\uparrow} - V_{\uparrow\leftarrow} < 0$ ($\Delta V > 0$) for $I < 0$ ($I > 0$) as spin polarization is generated in GaAs by the injection (extraction) of majority spins. As expected, the absolute value of ΔV increases with increasing bias, however the value of $\Delta R = \Delta V / I$ drops significantly at the same time, as is shown in the upper panel of Fig. 3(b). According to Eq.1. the latter could be attributed to the effect of the finite bias on $P_{inj}$ (we assume $P_{det}$ is not affected). At very low bias values the dependence saturates and for the lowest measured bias of 1 μA we can assume $P_{det} \approx P_{inj} = P$ and as a result we get $P_{inj} \approx 0.5$, which is comparable to values obtained in Esaki diodes experiments with spin-LEDs as detectors [28,29], as well as to the calculated ones [28]. From the bias dependence of ΔR we can then extract the bias dependence of $P_{inj}$, which we plot in the lower panel of Fig. 3(b). Its appearance for the reverse bias is very consistent with earlier reports [28,29], that explained a decrease in spin injection efficiency by increased contribution of minority spins to the tunnelling current at higher negative bias [28]. At forward bias a drop in $P_{inj}$ is initially slower than for the reverse case, and could be explained by inelastic tunnelling processes through forbidden states in the bandgap [16]. The pronounced contribution of such transitions to the total current is supported by the very small peak-to valley ratio measured at the dip in *I-V* curves. A dramatic drop in the in the $P_{inj}$ occurs at the bias of ~ 250 μA. As it coincides with the dip observed at the *I-V* characteristic of the injector, it could be explained by the fact that at this region the thermal current starts dominating over the tunnelling [16].

To summarize, we have demonstrated for the first time the successful injection of spin-polarized electrons from a ferromagnetic semiconductor into non-magnetic semiconductor and the subsequent detection by purely electrical means. The bias dependence of the obtained spin injection efficiency is consistent with the physics of tunnelling effects in Esaki diodes, employed as injecting and detecting contacts. Following the above discussion of this bias dependence it seems to be reasonable to raise the question whether and how optimising the performance of (Ga,Mn)As/GaAs Esaki diodes could improve already high value of generated spin polarization.

We thank J. Fabian for stimulating discussions. This work has been supported by the Deutscheforschungsgemeinschaft (DFG) through SFB689 project.

**Figure captions**

**Fig. 1.** (color online) (a) The scanning electron microscope (SEM) picture of the tilted device identical to the one measured. The size of all magnetic contacts (2 – 5) is 1 x 50 μm and the spacing between the centres of neighbouring contacts is 5 μm. The charge current flows only in the injector circuit whereas injected spins diffuse along *x* in either direction of the injecting contact. (b) The layers forming magnetic Esaki diode contacts and the schematic of the relative position of conduction band (cb) and valence band (vb) in the structure. Crossed areas indicate states occupied by electrons at T=0. **(c)** Current-voltage characteristic of the injector contact, typical also for other (Ga,Mn)As contacts. A dip in the current, characteristic for Esaki diodes, is clearly observed.

**Fig. 2.** (color online) (a) Non-local voltage measured at three different detector contacts vs. in-plane magnetic field $\mathbf{B}_x$. Arrows indicate sweep directions of respective curves. (b) dependence of the spin-valve signal $\Delta V$ on the injector – detector separation *L*. (c) The non-local voltage $V_{46}$ versus out-of-plane magnetic field $\mathbf{B}_z$. The raw data is shown. The background is fitted by a second order polynomial (blue dashed curve). For details see text. (d) Hanle curves obtained by subtracting the background signal from the non-local voltage measured at three different detectors. Solid curves are obtained from Eq. 2. with the fit parameters shown. All measurements at *T*=4.2K with $I_{21}$=−50 μA.

**Fig. 3.** (color online) (a) Non-local voltage $V_{36}$ versus in-plane field $\mathbf{B}_x$ (solid lines) and out-of-plane field $\mathbf{B}_z$ (symbols) for $I_{21}$=±5 μA. Grey solid bar indicates zero spin signal. (b) Upper panel: The value of the spin-signal $\Delta R=\Delta V/I$, measured at the contacts 3 and 4, versus bias current $I_{21}$. The region of very small bias currents is shown in the inset. Lower panel: The spin injection efficiency $P_{inj}$ (spin polarization of the injected current) versus bias current $I_{21}$. Inset: $P_{inj}$ vs. bias voltage $V_{in}$ across the injecting Esaki diode. $V_{in}$ was extracted from $V_{21}$ by subtracting the voltage drop across the GaAs channel. Solid lines are only guides for an eye.

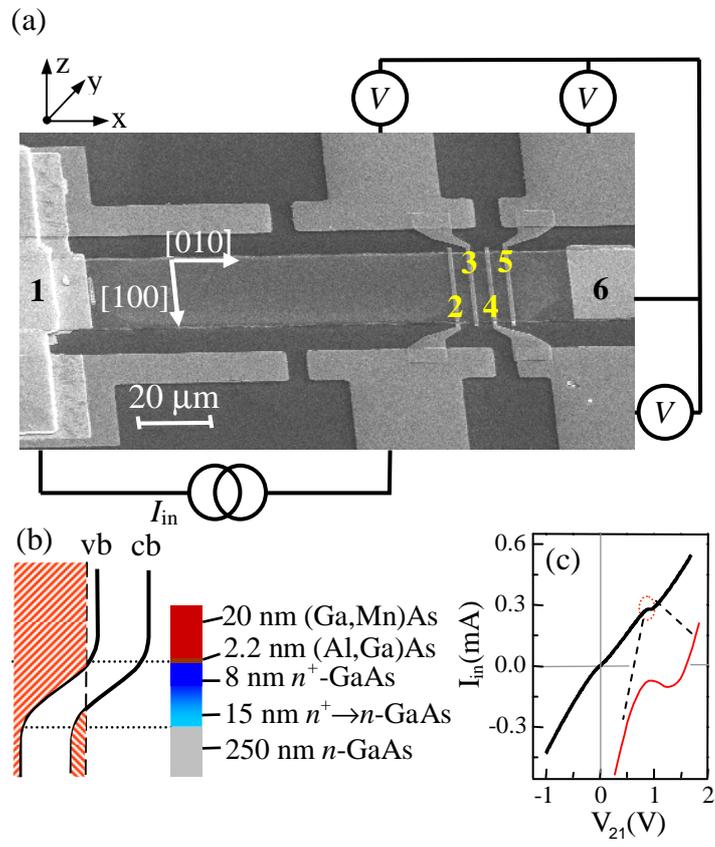

Figure 1

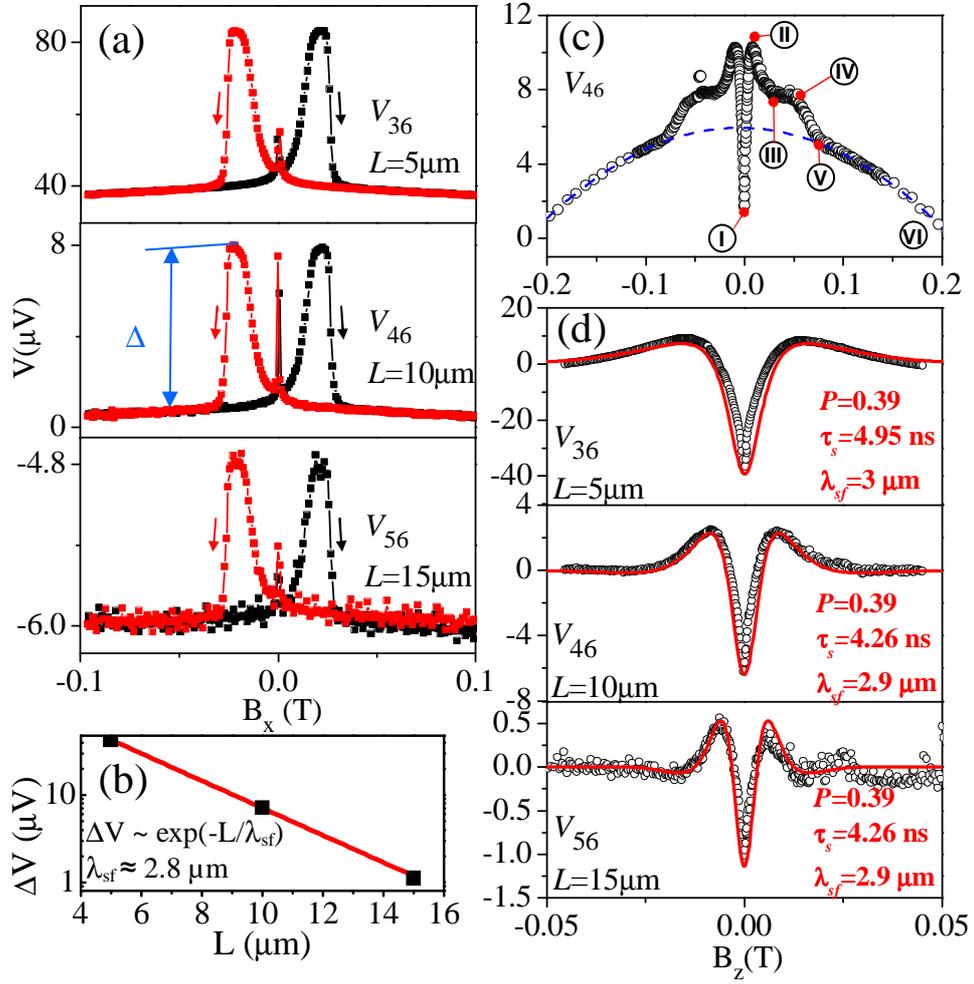

Figure 2

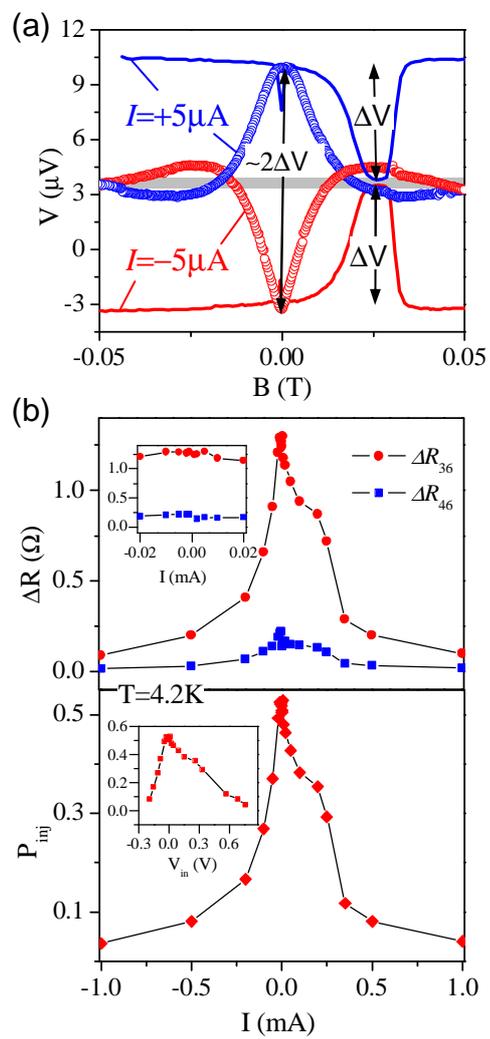

Figure 3